%% file: conf_rhopi.tex
\documentclass[11pt]{article}
\usepackage{graphicx}
\usepackage{epsf}
\usepackage{rotating}
\usepackage{epsfig}
\usepackage{supertabular}
\usepackage{subfigure}

\newcommand{\BABARPubYear}    {03}

\newcommand{\BABARConfNumber} {14}
\newcommand{\SLACPubNumber} {10078}

\input pubboard/babarsym
\input{Definitions}

\long\def\inst#1{\par\nobreak\kern 4pt\nobreak
    {\it #1}\par\vskip 10pt plus 3pt minus 3pt}

\begin{document}
{\pagestyle{empty}

\begin{flushright}
\babar-CONF-\BABARPubYear/\BABARConfNumber \\
SLAC-PUB-\SLACPubNumber \\
July 2003 \\
\end{flushright}

\par\vskip 3cm

\begin{center}
\Large \bf \boldmath 
	Measurement of Branching Fractions and  
	\CP-Violating \\
	Charge Asymmetries in $B^+\rar\rho^+\pi^0$ 
	and $B^+\rar\rho^0\pi^+$ decays, \\
	and Search for $B^0\rar\rho^0\pi^0$
\end{center}
\bigskip

\begin{center}
\large The \babar\ Collaboration\\
\mbox{ }\\
\today
\end{center}
\bigskip \bigskip

\begin{center}
\large \bf Abstract
\end{center}

\noindent
We present preliminary measurements of branching fractions and  
\CP-violating charge asymmetries in \B-meson decays to $\rho\pi$.
The data sample comprises 89 million $\FourS \to B\Bbar$
decays collected with the \babar\ detector at the \pep2 
asymmetric-energy $B$~Factory at SLAC.  We find the charge-averaged
branching fractions
${\cal B}( B^{+}\rightarrow \rho^{+}\pi^0) = (11.0 \pm 1.9{\rm (stat.)}
\pm 1.9{\rm (syst.)})\times
10^{-6}$ and ${\cal B}( B^{+} \rightarrow \rho^0 \pi^{+}) = (9.3 \pm 
1.0{\rm (stat.)} \pm 0.8{\rm (syst.)}) \times 10^{-6}$; 
we set a $90\%$ confidence-level upper limit
of ${\cal B}( \B^0 \rightarrow \rho^0\pi^0) < 2.5 \times 10^{-6}$.  
We measure the \CP-violating charge asymmetries 
$A_{CP}^{\rho^{+}\pi^0} = 0.23 \pm 0.16{\rm (stat.)} \pm 0.06{\rm (syst.)}$ and 
$A_{CP}^{\rho^0\pi^{+}} = -0.17 \pm 0.11{\rm (stat.)} \pm 0.02{\rm (syst.)}$.  
\vfill
\begin{center}
Contributed to the International Europhysics Conference on High Energy Physics\\ 
EPS (July 17 - July 23 2003) in Aachen, Germany.
\end{center}

\vspace{1.0cm}
\begin{center}
{\em Stanford Linear Accelerator Center, Stanford University, 
Stanford, CA 94309} \\ \vspace{0.1cm}\hrule\vspace{0.1cm}
Work supported in part by U.S. Department of Energy contract DE-AC03-76SF00515.
\end{center}

\newpage
} 

\input pubboard/authors_eps2003.tex

\input{Introduction}

\input{Detector}

\input{Discriminating}

\input{Selection}

\input{BBackground}

\input{MLfit}

\input{Results}

\input {Systematics}

\input{Summary}

\section{Acknowledgments}
\label{sec:Acknowledgments}

\input pubboard/acknowledgements

\input{Biblio}
\end{document}

%% file: Definitions.tex
\newcommand{\bit}{\begin{itemize}}
\newcommand{\eit}{\end{itemize}}
\newcommand{\beq}{\begin{equation}}
\newcommand{\eeq}{\end{equation}}
\newcommand{\beqn}{\begin{eqnarray}}
\newcommand{\eeqn}{\end{eqnarray}}
\newcommand{\beqns}{\begin{eqnarray*}}
\newcommand{\eeqns}{\end{eqnarray*}}

\def\rar{\rightarrow}

\def\btorp{\Bz \rar \rho\pi}

\def\bchtorpch{B^{+} \rar \rho^0 \pi^{+}}
\def\bchtorchp{B^{+} \rar \rho^{+} \pi^0}
\def\btorp{B^{0} \rar \rho^0 \pi^0}

\def\rhopi{\rho\pi}

\def\hel{{\theta_{\pi}}}

\def\cshelrho{{\rm cos}\,\hel}

\def\de{\Delta E}
\def\mes{\ensuremath{m_{\rm ES}}}

\def\thetaC{\theta_C}

\def\cat{k}

\def\MeVc2{${\rm MeV}/c^2$}
\def\GeVc2{${\rm GeV}/c^2$}

\def\BR{{\cal BR}}

\def\NN{{\rm NN}}

\def\scf{{\rm scf}}

\def\ea{{\em et al.}}

\def\mes {\ensuremath{m_{ES}}\xspace}

\def\Dbar{\ensuremath{\overline{D}}\xspace}

\def\dt {\ensuremath{\Delta t}}

%% file: pubboard/authors_eps2003.tex
\begin{center}
\small

The \babar\ Collaboration,
\bigskip

%
B.~Aubert,
R.~Barate,
D.~Boutigny,
J.-M.~Gaillard,
A.~Hicheur,
Y.~Karyotakis,
J.~P.~Lees,
P.~Robbe,
V.~Tisserand,
A.~Zghiche
\inst{Laboratoire de Physique des Particules, F-74941 Annecy-le-Vieux, France }
% ...

\end{center}\newpage

%% file: Introduction.tex
\section{Introduction}
\label{sec:Introduction}

The study of \B meson decays into charmless hadronic final states 
plays an important role in the understanding of the phenomenon of \CP 
violation in the \B system. Recently, the \babar\  experiment has 
performed a search for \CP-violating asymmetries in neutral 
$B$ decays to $\rho^\pm\pi^\mp$~\cite{bib:350PRL}, where the 
mixing-induced \CP asymmetry is related to the angle 
$\alpha \equiv \arg\left[-V_{td}^{}V_{tb}^{*}/V_{ud}^{}V_{ub}^{*}\right]$ 
of the Unitarity Triangle. However, in contrast to the theoretically 
clean determination of $\sin2\beta$ in the decay to charmonium 
such as $\Bz\to\jpsi\KS$~\cite{babarsin2b,bellesin2b},
the extraction of $\alpha$ from $\rho^\pm\pi^\mp$ is complicated 
by the interference of decay amplitudes with different weak phases.
Various strategies to overcome this problem have been proposed 
in the literature.  One such method is an SU(2) isospin analysis 
of the $\rho\pi$ final states~\cite{bib:Nir}. In the limit of 
isospin symmetry, the five decay amplitudes $B^0\rar\rho^+\pi^-$,
$B^0\rar\rho^-\pi^+$, $B^0\rar\rho^0\pi^0$, $B^+\rar\rho^+\pi^0$ and 
$B^+\rar\rho^0\pi^+$ form a pentagon in the complex plane. 
Combining measurements of all the decay rates, mixing-induced and 
direct \CP asymmetries in the neutral \B modes, as well as charge 
asymmetries in the charged \B modes, allows a determination of the 
phase $\alpha$ that is free of hadronic uncertainties.

In this letter, we present preliminary measurements of the branching 
fractions of the decay modes\footnote
{
	If not stated otherwise, charge conjugation is implied 
	throughout this document.
} $B^+\rar\rho^0\pi^+$, $B^+\rar\rho^+\pi^0$, 
and perform a search for the decay $B^0\rar\rho^0\pi^0$. For the 
charged modes we also present measurements of the \CP-violating 
charge asymmetry $A_{CP}$, defined by
\beq
\label{equ:Acp}   
	A_{CP} \equiv 
	\frac{\Gamma(B^- \rightarrow f) \,-\, \Gamma(B^+ \rightarrow \bar{f})}
	 {\Gamma(B^- \rightarrow f) \,+\, \Gamma(B^+ \rightarrow \bar{f})}
\eeq
where $\Gamma(B^- \rightarrow f)$ and $\Gamma(B^+ \rightarrow \bar{f})$ 
are the $\Bub$ and $\Bu$ decay rates,
respectively. A non-zero $A_{CP}$ requires the presence of at least
two amplitudes with different weak and strong phases.

The measurements of the branching fractions and charge asymmetries use
events collected by the $\babar$ detector at the \pep2 asymmetric-energy 
$B$~Factory. The sample used for the charged modes consists of 
$88.5\times10^6$ \BB\ pairs collected at the $\FourS$ 
resonance (``on-resonance''), while the $B^0\rar\rho^0\pi^0$ 
search uses a slightly larger sample of $88.9\times 10^6$ \BB\ pairs. 
We use an integrated luminosity of 9.6\invfb collected 
approximately 40\mev below the  $\FourS$ ("off-resonance") for 
background studies. We determine the yields and charge asymmetries
using a maximum likelihood (ML) fit.

%% file: Detector.tex
\section{The \babar~Detector}
\label{sec:detector}

A detailed description of the \babar~detector can be found in
Ref.~\cite{bib:babarNim}.  Charged-particle momenta are measured in a
tracking system consisting of a $5$-layer double-sided silicon vertex
tracker (SVT) and a $40$-layer drift chamber (DCH) filled with a gas
mixture based on helium and isobutane. The SVT and DCH operate within
a 1.5\,T superconducting solenoidal magnet. The typical decay vertex
resolution is around $65\mum$ along the beam direction for the
fully-reconstructed~$B_{\rho\pi}$ (referred to as signal $B$ hereafter),
and around $185\mum$ 
for the inclusively-reconstructed rest of the event.  Photons are 
detected in an electromagnetic calorimeter (EMC) consisting of $6580$ 
CsI(Tl) crystals arranged in barrel and forward end-cap sub-detectors.  
The average $\piz$ mass resolution is $7 \mevcc$.  The flux 
return for the solenoid is composed of multiple layers of iron and 
resistive plate chambers for the identification of muons and long-lived 
neutral hadrons.  Tracks are identified as 
pions, kaons or protons by the Cherenkov angle $\thetaC$ measured with a 
detector of internally reflected Cherenkov light (DIRC). The typical 
separation between pions and kaons  varies from $8\sigma$ at 
$2\gevc$ to $2.5\sigma$ at $4\gevc$, where $\sigma$ is the average 
$\thetaC$ resolution. Lower momentum kaons are identified with a 
combination of $\thetaC$ (for momenta down to $0.7\gevc$) and 
measurements of energy loss, $dE/dx$, in the DCH and SVT.

%% file: Discriminating.tex
\section{Discriminating Variables}
\label{sec:Discriminating}

To reject background from continuum $e^+e^-\to q\bar{q}$ 
($q = u,d,s,c$) events and other $B$ decays, we use the 
following discriminating variables:
\begin{itemize}

\item 	$\mes$: the beam-energy-substituted mass is defined by
	\beq
	\mes = \sqrt{(s/2 + {\mathbf {p}}_i\cdot {\mathbf {p}}_B)^2/E_i^2 
			- {\mathbf {p}}_B^2}~,
	\eeq
	where $s$ is the square of the center-of-mass (CM) energy, 
	$E_i$ and ${\mathbf {p}}_i$ are the total energy and 
	three-momentum of the initial \epem\ state in the
	laboratory frame, and ${\mathbf {p}}_B$ is the three-momentum 
	of the $B$ candidate in the same frame. Signal events 
	populate the $\mes$ region near the $B$ mass. Background events
        usually have a wider or even flat $\mes$ distribution.

\item 	$\de$: the difference between the reconstructed energy of 
	the signal $B$ candidate and the beam energy in 
	the CM frame. The $\de$ distribution for signal events 
        peaks around zero. Backgrounds from other $B$-meson decays
	peak at different $\de$ depending on the number of
	charged and neutral particles in the decay. Two-body and 
	four-body decays populate the positive and negative regions 
	of $\de$, respectively, while three-body decays have 
	$\de$ values near zero. The $\de$ is shifted if
        a misidentified charged track is used to reconstruct the 
        signal $B$.

\item 	Output of a multivariate analyzer : 
	the dominant background for charmless $B$ decays 
	are continuum events. To enhance discrimination between 
	signal and continuum, three neural networks (NN), one for 
	each signal mode, are constructed. Each NN is trained 
	with off-resonance data and simulated events of its 
	corresponding signal. All NNs use: 
	the reconstructed $\rho$ mass; 
	$\cshelrho$, where $\hel$ is the angle between the momentum 
        of the $\pi^+$($\pi^0$) from the $\rho^0$($\rho^+$) decay
	and the $B$ momentum in the $\rho$ rest frame; 
	the event shape variables $L_0 =\sum_{i} p^*_i$ and
	$L_2 = \sum_{i} p^*_i \times|\cos(\theta^*_{T_B,i})|^2$, 
	where $p^*_i$ is the CM momentum of the track or neutral object 
	$i$, belonging to the rest of the event, and $\theta^*_{T_B,i}$ 
	is the angle between the momentum of track $i$ and the $B$ 
	thrust axis $T_B$ in the CM frame; 
	$|\cos(\theta_{B,z})|$, the cosine of the angle
	between the \B momentum and the $z$ axis
	(along the beam direction) in the CM frame, and 
	$|\cos(\theta_{T_B,z})|$, the cosine of the angle between 
	the \B thrust axis and the $z$ axis in the CM frame.

\end{itemize}

%% file: Selection.tex
\section{Event Selection and Reconstruction}
\label{sec:Selection_All_channels}

Signal \B candidates are reconstructed from combinations of three-pion 
final states that must be $\pi^{+}\pi^0\pi^0$ for 
$B^+\rar\rho^+\pi^0$, $\pi^{+}\pi^-\pi^+$ for $B^+\rar\rho^0\pi^+$,
and $\pi^{+}\pi^{-}\pi^0$ for $B^0\rar\rho^0\pi^0$.  The charged 
tracks must be inconsistent with being an electron based on $dE/dx$ 
measurements, shower shape criteria in the EMC, and the ratio of 
shower energy and track momentum. Charged tracks must also be 
inconsistent with being a kaon or a proton. The photon candidates
used to form a $\pi^0$ must have an energy greater than $50\mev$ 
in the laboratory frame, and exhibit a lateral shower profile of 
energy deposition in the EMC consistent with an electromagnetic shower.
The invariant mass $m(\gamma\gamma)$ of the photon pair 
must satisfy $0.11<m(\gamma\gamma)<0.16\gevcc$. 

Two of the three final state pions are used to form a charged or a 
neutral $\rho$ candidate (the third pion is referred to as the bachelor 
pion hereafter). For the $\bchtorchp$ mode, the $\rho^{+}$ candidate 
is reconstructed from the positively charged track and one of the two 
$\pi^0$ candidates. For the $\bchtorpch$ and $\btorp$ modes, the 
$\rho^{0}$ candidate is reconstructed from two oppositely-charged 
tracks. The mass of the $\rho$ candidates must satisfy
$0.4<m(\pi^{+}\pi^0)<1.3\gevcc$ for $\bchtorchp$,
$0.53<m(\pi^+\pi^-) < 0.9\gevcc$ for $\bchtorpch$, and 
$0.4<m(\pi^+\pi^-) < 0.9\gevcc$ for $\btorp$. 
The $\bchtorpch$ mass cut has been tightened to remove \B-related 
backgrounds such as $B^{+} \rightarrow f_0(980)\pi^{+}(K^{+})$, 
$\KS \pi^{+}$. The $\btorp$ mass cut is tight to remove 
$B^0\rar f_0(980)\pi^0$. For the $\btorp$ mode, the corners
of the Dalitz plot where $\rho^0\pi^0$ interferes with the dominant
$\rho^{\pm}\pi^{\mp}$ are removed by requiring that the invariant 
masses $m(\pi^+\pi^0)$ and $m(\pi^-\pi^0)$ be greater than $1.3\gevcc$.  
To take advantage of the helicity structure of $\B \rightarrow \rho\pi$
decays, we require $|\cos\theta_{\pi}|>0.25$.

To reject background from $B^+\rightarrow \rho^0 K^+$ and
$B^0\rightarrow \rho^{-}K^{+}$ decays, only $\rho^0\pi^+$
candidates with bachelor tracks within the geometrical 
acceptance of the DIRC are considered. The number of photons 
in the DIRC that are associated with the bachelor track must 
not be less than 5.

For the $\bchtorchp$ and $\btorp$ modes, the invariant mass of 
either track and the $\piz$ must be less than $5.14\gevcc$ to 
reject two-body \B background.

\B candidates are required to satisfy kinematic fit-region cuts. 
For $\bchtorchp$
decays, candidates must satisfy $5.20 < \mes < 5.29\gevcc$ and 
$-0.15<\de<0.10\gev$. $\bchtorpch$ candidates must satisfy
$5.23 < \mes < 5.29\gevcc$ and $-0.05< \de < 0.05\gev$. The
$\bchtorpch$ analysis benefits from a better $\de$ resolution 
due to the absence of neutral pions; the tight cut on $\de$ helps 
to remove four-body \B background more effectively. $\btorp$
candidates must satisfy $5.23<\mes<5.29\gevcc$ and 
$-0.15<\de< 0.10\gev$.

For the $\bchtorpch$ mode, we remove background from 
charmed decays $B\rightarrow \Dzb X$, $\Dzb\rightarrow K^{+}\pi^{-}$ 
or $\pi^{+}\pi^{-}$, by requiring that all pairs of oppositely-charged 
tracks have invariant masses either smaller than $1.844\gevcc$ 
or greater than $1.884\gevcc$, assuming both kaon and pion hypotheses 
for the positively-charged track.

The final samples of signal candidates are selected with a cut
on the NN outputs for all three decay modes. For example, the NN 
cut for the $\bchtorpch$ decay mode retains 85\%(11\%) of the signal
(continuum) events.

In each event, final state particles other than the three pions 
that form the signal $B$ meson are assumed to belong to the other 
$B$ meson. These particles are used to tag the flavor of the other 
$B$ meson and to inclusively-reconstruct its vertex for decay time 
determination. In this letter, this other $B$ is referred to as 
$B_{\rm tag}$.

For the $\btorp$ mode, we use the proper decay time as a discriminating 
variable in the ML fit. The time difference $\dt$ is obtained from 
the measured distance between the $z$ positions (along the beam direction) 
of the $\B_{\rho^0\piz}$ and $\B_{\rm tag}$ decay vertices, and the known 
boost of the \epem\ system. The vertex of the $\B_{\rm tag}$ is 
reconstructed from all tracks in the event except those
from the $\B_{\rho^0\piz}$, and an iterative procedure~\cite{babarsin2b}
is used to remove tracks with a large contribution to the 
vertex~$\chi^2$. An additional constraint is obtained from the 
three-momentum and vertex position of the $\B_{\rho^0\piz}$ 
candidate, and the average \epem\ interaction point and boost. 
We require $|\dt| < 20\ps$ and $\sigma(\Delta t)<2.5\ps$, 
where $\sigma(\Delta t)$ is the error on $\dt$ estimated on an
event-by-event basis.

Approximately $33\%$ ($7\%$ and $8\%$) of the events from signal
Monte Carlo (MC) simulation have more than one candidate satisfying the 
selection in the $\bchtorchp$ ($\bchtorpch$ and $\btorp$) decay mode.
In this case, we choose the candidate with the reconstructed 
$\rho$ invariant mass closest to the nominal $\rho$ mass~\cite{PDG}. 
Any chosen candidate from a signal event that is reconstructed
from one or more wrong particles (charged or neutral) is hereafter
referred to as misreconstructed signal.

The signal selection efficiencies are determined by applying the
selection criteria to MC. Efficiencies
(including misreconstructed signal) of the three decay modes
are summarized in Table~\ref{tab:effAll}. Also given in 
Table~\ref{tab:effAll} are the fractions of misreconstructed events in the 
selected signal samples. Misreconstruction is mostly due to 
$\rho$ candidates that include a random low-momentum pion.
Some of the misreconstructed events are assigned an incorrect \B 
charge in charged \B decays. The total number of events that 
enter the likelihood fits are given in the last row of
Table~\ref{tab:effAll}.  

\begin{table}[t]
\begin{center}
\caption{\em Signal efficiencies ($\epsilon$), fractions of 
	misreconstructed signal events ($f_\scf$), and fractions 
        of misreconstructed signal events with wrong \B
	candidate charge ($\omega_Q$) in selected 
	MC-simulated  events. The last row gives the numbers of
       	selected on-resonance events entering the maximum likelihood
	fits. }
\label{tab:effAll}
\vspace{0.3cm}
\setlength{\tabcolsep}{1.1pc}
\begin{tabular}{lccc} \hline\hline
&&&\\[-0.35cm]
\rule[-2.3mm]{0mm}{5mm}
                   & $\bchtorchp$  & $\bchtorpch$  & $\btorp$ \\\hline
&&&\\[-0.35cm]
\rule[-2.3mm]{0mm}{5mm}
$\epsilon$ $[\%]$  & $17.6\pm 0.1$ & $29.0\pm 0.1$ & $20.2\pm 0.1$  \\
\rule[-2.3mm]{0mm}{5mm}
$f_\scf$ $[\%]$    & $38.6\pm 0.2$ & $7.1\pm 0.1$  & $9.4\pm 0.2$ \\
\rule[-2.3mm]{0mm}{5mm}
$\omega_Q$ $[\%]$  & $8.1\pm 0.1$  & $1.6\pm 0.1$  & - \\
\rule[-2.3mm]{0mm}{5mm}
Selected events    & 13104         & 8498          & 6648 \\\hline\hline
\end{tabular}
\end{center}
\end{table}

While flavor information of the $B_{\rm tag}$ is not explicitly used in 
our analyses, event 
properties used to categorize flavor-tagged events can be exploited to 
distinguish signal and (mainly) continuum background because they populate
among tagging categories in dramatically different ways. Five mutually 
exclusive tagging categories are defined: {\tt Lepton}, {\tt Kaon}, 
{\tt NT1}, {\tt NT2}, and {\tt Untagged}. They are determined by a
tagging algorithm~\cite{babarsin2b} relying on the correlation between 
the flavor of the $b$~quark and the charge of the remaining tracks in 
the event, after removal of the tracks from the $B\to\rho \pi$.
The fractions of signal events in each tagging category are measured 
from data using a control sample of fully-reconstructed $B$ 
decays~\cite{babarsin2b}. Tagging fractions for \B backgrounds 
are taken from Monte Carlo. Separate continuum background yields 
for each category are free to vary in the maximum likelihood fit.
Using tagging categories, we decrease the statistical error in the signal 
yield by about $10\%$.

%% file: BBackground.tex
\section{\boldmath$B$-related Backgrounds}
\label{sec:BBackground}

We use MC simulation to study the cross-feed from 
two-body, three-body and four-body charmless $B$~decays, as well as
from inclusive $b \rightarrow c$ decays.
The branching fractions of 
unmeasured decay channels are estimated within conservative error ranges. 
The charmless modes are grouped into eighteen ($\bchtorchp$), fifteen 
($\bchtorpch$) and seventeen ($\btorp$) classes with similar kinematic 
and topological properties. Two additional classes account for the 
neutral and charged $b \to c$ decays. For each of the background 
classes, a component is introduced into the ML fit, with a fixed 
number of events. 
Contributions to the systematic error by each B background mode is obtained
by varying its yield by one standard deviation. For unknown modes with 
only estimated branching ratios, the uncertainties are divided by 
$\sqrt{3}$ and taken as the standard deviations.
Tables~\ref{tab:BbkgClass_rchp}, \ref{tab:BbkgClass_rpch} and 
\ref{tab:BbkgClasse_rp} summarize the dominant \B background modes 
to $\bchtorchp$, $\bchtorpch$ and $\btorp$, respectively. 

\begin{table}[thp]
\begin{center}
\caption{\em \B background modes considered in the 
	$\Bp\to\rho^{+}\pi^0$ maximum likelihood fit. The second 
	column gives the branching fractions used (estimated branching 
	fractions are indicated by an asterisk), and the third column 
	quotes the expected number of events entering into the sample,
        scaled to $81.5\invfb$ ($88.5\times10^6 ~\FourS \to \B\Bbar$ decays).
	The labels ``long.'' and ``tran.'' refer to the longitudinal
	and transverse polarization of the final states, respectively, 
	in \B decays into vector-vector mesons. $K^{(**)}_X$ refers to the
        higher kaonic resonances including $K_0^*(1430)$, 
	$K_2^*(1430)$ and $K^*(1680)$. $X_c$ indicates inclusive charmed 
        decays.}
\label{tab:BbkgClass_rchp}
\vspace{0.3cm}
\setlength{\tabcolsep}{1.6pc}
\begin{tabular}{lcc} \hline\hline
&&\\[-0.35cm]
\rule[-2.3mm]{0mm}{5mm}
 Mode & BR $[10^{-6}]$ & N$_{\rm exp}$ \\
\hline
&&\\[-0.35cm]
\rule[-2.3mm]{0mm}{5mm}
  $B^0  \rightarrow  \pi^0\pi^0$                 & $1.6\pm1.6$   &  $4.8\pm4.8$ \\
\rule[-2.3mm]{0mm}{5mm}
  $B^+  \rightarrow  \pi^+\pi^0$                 & $5.2\pm0.8$   &  $5.6\pm0.9$ \\
\rule[-2.3mm]{0mm}{5mm}
  $B^+  \rightarrow  K^+\pi^0$                   & $12.7\pm1.2$   &  $4.9\pm0.5$ \\

\rule[-2.3mm]{0mm}{5mm}
  $B^0 \rightarrow  \rho^\pm \pi^\mp$            & $22.6\pm2.8$   & $44.6\pm5.6$ \\
\rule[-2.3mm]{0mm}{5mm}
  $B^0 \rightarrow  \rho^- K^+$              & $7.3 \pm 1.8 $   & $2.5\pm0.6$ \\
\rule[-2.3mm]{0mm}{5mm}
  $B^+  \rightarrow  \KS(\pi^0\pi^0)\pi^+$ & $4.1\pm0.4$      & $5.4\pm0.5$ \\
\rule[-2.3mm]{0mm}{5mm}
  $B^0 \rightarrow   K^{*0}(K^+\pi^-)\pi^0$      &  $8.7\pm5.0$  &  $3.8\pm2.2$ \\
\rule[-2.3mm]{0mm}{5mm}
  $B^0  \rightarrow  K^{*0}(\KS\pi^0)\pi^0$      &  $7.5\pm5.0$   &  $3.9\pm2.6$ \\
\rule[-2.3mm]{0mm}{5mm}
  $B^+  \rightarrow  K^{*+}(K^+\pi^0)\pi^0$      &  $4.4\pm2.5$   &  $6.1\pm3.5$ \\
\rule[-2.3mm]{0mm}{5mm}
  $B^+  \rightarrow  \rho^+\gamma$               & $2.3\pm2.3$  &  $1.3\pm1.3$ \\

\rule[-2.3mm]{0mm}{5mm}
  $B^0 \rightarrow  \rho^+\rho^-$ long.          & $40.0^{+50*}_{-35}$ & $64.8^{+81}_{-57}$ \\
\rule[-2.3mm]{0mm}{5mm}
  $B^0 \rightarrow  \rho^+\rho^-$ tran.         & $40.0^{+50*}_{-35}$  & $1.6^{+2.0}_{-1.4}$ \\
\rule[-2.3mm]{0mm}{5mm}
  $B^+ \rightarrow  \rho^+\rho^0$ long.        & $30.1^{+8.3}_{-9.9}$  & $17.0^{+4.3}_{-5.2}$ \\
\rule[-2.3mm]{0mm}{5mm}
  $B^+  \rightarrow  a_{1}^+\pi^0$               & $35\pm35^*$    & $25.0\pm25.0$ \\

\rule[-2.3mm]{0mm}{5mm}
  $B^+  \rightarrow  (K^{(**)}_X\pi)^+$          & $40\pm26^*$    & $15\pm9.8$ \\
\rule[-2.3mm]{0mm}{5mm}
  $B^+  \rightarrow  (K^{(**)}_X\rho)^+$         & $20\pm20^*$    & $1.2\pm1.2$ \\
\rule[-2.3mm]{0mm}{5mm}
  $B^0  \rightarrow  (K^{(**)}_X\pi)^0$          & $72\pm54^*$    & $23\pm17.3$ \\
\rule[-2.3mm]{0mm}{5mm}
  $B^0  \rightarrow  (K^{(**)}_X\rho)^0$         & $20\pm20^*$    & $6.0\pm6.0$ \\
\hline
&&\\[-0.35cm]
\rule[-2.3mm]{0mm}{5mm}
 Total charmless background & & $236.5\pm 79.3$ \\ \hline
&&\\[-0.35cm]
\rule[-2.3mm]{0mm}{5mm}
  $B^0\to X_c^0$                    &       --       &   $72.0\pm16.0$
\\
\rule[-2.3mm]{0mm}{5mm}
  $B^+\to X_c^+$                    &       --       &  $133.0\pm30.0$
\\
\hline
&&\\[-0.35cm]
\rule[-2.3mm]{0mm}{5mm}
Total \B background & & $442 \pm 86$ \\
\hline\hline
\end{tabular}
\end{center}
\end{table}

\begin{table}[htb]
\begin{center}
\caption{\em \B background modes considered in the 
	$\Bp\to\rho^0\pi^{+}$ maximum likelihood fit. The second 
	column gives the branching fractions used (estimated branching 
	fractions are indicated by an asterisk), and the third column 
	quotes the expected number of events entering into the sample, 
        scaled to $81.5\invfb$ ($88.5\times10^6 ~\FourS \to \B\Bbar$ 
        decays). }
\label{tab:BbkgClass_rpch}
\vspace{0.3cm}
\setlength{\tabcolsep}{1.6pc}
\begin{tabular}{lcc}
\hline\hline
&&\\[-0.35cm]
\rule[-2.3mm]{0mm}{5mm}
Mode & BR $[10^{-6}]$ & N$_{\rm exp}$ \\\hline
&&\\[-0.35cm]
\rule[-2.3mm]{0mm}{5mm}
$B^0 \rightarrow \rho^{\pm}\pi^{\mp}$ & $22.6 \pm 2.8$ & $29.3 \pm 5.2$ \\
\rule[-2.3mm]{0mm}{5mm}
$B^0 \rightarrow \rho^-K^+$ & $7.3 \pm 1.8$ & $ 1.1 \pm 0.3$ \\
\rule[-2.3mm]{0mm}{5mm}
$B^+ \rightarrow \rho^0K^+$ & $3.9 \pm 1.2$  & $4.9 \pm1.5$\\
\rule[-2.3mm]{0mm}{5mm}
$B^+ \rightarrow f_0(980)K^+$ & $11.7 \pm 4.0$ & $1.1 \pm 0.4$ \\
\rule[-2.3mm]{0mm}{5mm}
$B^+ \rightarrow \KS(\pi^+\pi^-)\pi^+$ & $9.0 \pm 0.9$ & $ 5.3 \pm 0.5$ \\
\rule[-2.3mm]{0mm}{5mm}
$B^0 \rightarrow K^{*+}(K^+\pi^0)\pi^-$ & $8.7 \pm 5.0$ & $1.4 \pm 0.8$ \\
\rule[-2.3mm]{0mm}{5mm}
$B^+ \rightarrow K^{*0}(K^+\pi^-)\pi^+$ & $10.3 \pm 2.6$ & $11.1\pm 2.8$ \\
\rule[-2.3mm]{0mm}{5mm}
$B^+\rightarrow \pi^+\omega(\pi^+\pi^-) $  & $0.14\pm0.04$ & $3.6 \pm 1.0$ \\
\rule[-2.3mm]{0mm}{5mm}
$B^0 \rightarrow \rho^+\rho^-$ long. & $ 40^{+50*}_{-35}$ & $6.3^{+7.8}_{-5.5}$\\
\rule[-2.3mm]{0mm}{5mm}
$B^0 \rightarrow \rho^0\rho^0$ long. & $ 3.5 \pm 3.5^*$ & $1.7 \pm 1.7$ \\
\rule[-2.3mm]{0mm}{5mm}
$B^+\rightarrow\rho^+\rho^0$ long. & $30.1^{+8.3}_{-9.9}$ & $7.9^{+2.2}_{-2.6}$ \\
\rule[-2.3mm]{0mm}{5mm}
$B^+\rightarrow \eta^{\prime}(\rho^0\gamma)\pi^+$ & $ 3.0 \pm 2.0^* $ &$2.2 \pm 1.5$\\
\rule[-2.3mm]{0mm}{5mm}
$B^0 \rightarrow a_1^+\pi^-$ & $35 \pm 35^*$ & $5.3 \pm 5.3$\\
\rule[-2.3mm]{0mm}{5mm}
$B^+ \rightarrow (K^{(**)}_{X} \pi)^+$ & $40\pm26^*$ & $2.9\pm 1.9$  \\
\rule[-2.3mm]{0mm}{5mm}
$B^0 \rightarrow (K^{(**)}_{X} \pi)^0$ & $72\pm54^*$ & $7.4\pm 5.5$  \\\hline
&&\\[-0.35cm]
\rule[-2.3mm]{0mm}{5mm}
Total charmless background & - & 91.5$\pm$11.3 \\\hline
&&\\[-0.35cm]
\rule[-2.3mm]{0mm}{5mm}
$B^0 \to X_c^0$ & -           & $19.2 \pm 5.8 $ \\
\rule[-2.3mm]{0mm}{5mm}
$B^+ \to X_c^+$ & -           & $54.1 \pm 13.1 $ \\\hline
&&\\[-0.35cm]
\rule[-2.3mm]{0mm}{5mm}
Total \B background & -           & $165 \pm 19$ \\\hline\hline
\end{tabular}
\end{center}
\end{table}

\begin{table}[thp]
\begin{center}
\caption{\em \B background modes considered in the 
	$\Bz\to\rho^{0}\pi^0$ maximum likelihood fit. The second 
	column gives the branching fractions used (estimated branching 
	fractions are indicated by an asterisk), and the third column 
	quotes the expected number of events entering into the sample,
        scaled to $81.9\invfb$ ($88.9\times10^6 ~\FourS \to \B\Bbar$ 
        decays). }
\label{tab:BbkgClasse_rp}
\vspace{0.3cm}
\setlength{\tabcolsep}{1.6pc}
\begin{tabular}{lcc}
\hline\hline
&&\\[-0.35cm]
\rule[-2.3mm]{0mm}{5mm}
Mode & BR $[10^{-6}]$ & N$_{exp}$ \\ \hline
&&\\[-0.35cm]
\rule[-2.3mm]{0mm}{5mm}
 $ B^+ \rightarrow \pi^+\pi^0$ & 5.2 $\pm$ 0.8   & 1.4  $\pm$ 0.2 \\
\rule[-2.3mm]{0mm}{5mm}
 $ B^+ \rightarrow  K^+\pi^0$   & 12.7 $\pm$ 1.2 & 0.9 $\pm$ 0.1 \\

\rule[-2.3mm]{0mm}{5mm}
 $ B^0 \rightarrow \rho^\pm\pi^\mp$ & 22.6 $\pm$ 2.8 & 16.9 $\pm$ 1.4  \\
\rule[-2.3mm]{0mm}{5mm}
 $ B^0 \rightarrow \rho^- K^+ $ & 7.3  $\pm$ 1.8 & 0.7 $\pm$ 0.2 \\
\rule[-2.3mm]{0mm}{5mm}
 $ B^+ \rightarrow \rho^+\pi^0$ & 15.0 $^{+15}_{-10}$* & 15.8 $^{+15.8}_{-10.5}$  \\
\rule[-2.3mm]{0mm}{5mm}
 $ B^0 \rightarrow \KS(\pi^+\pi^-)\pi^0$  & 3.5 $\pm$ 0.5 & 5.0 $\pm$ 0.7  \\
\rule[-2.3mm]{0mm}{5mm}
 $ B^0 \rightarrow f_0(980) \pi^0$ & 0.0 $\pm$ 3.0* & 0.0 $^{+3.1}_{-0.0}$ \\
\rule[-2.3mm]{0mm}{5mm}
 $ B^0 \rightarrow \pi^+ \pi^- \pi^0$ (non-res.)  & 0.0 $\pm$ 5.0* & 0.0 $^{+4.1}_{-0.0}$ \\
\rule[-2.3mm]{0mm}{5mm}
 $ B^0 \rightarrow K^{*0}(K^+\pi^-)\pi^0$ & 8.7 $\pm$ 5.0* & 13.6 $\pm$ 7.8 \\
\rule[-2.3mm]{0mm}{5mm}
 $ B^0 \rightarrow K^{*0}((K\pi)^0)\gamma$ & 40.2 $\pm$ 2.7 & 1.7 $\pm$ 0.1 \\

\rule[-2.3mm]{0mm}{5mm}
 $ B^0 \rightarrow \rho^+\rho^-$ long. & 40.0 $^{+50}_{-35}$* & 10.9 $^{+13.6}_{-9.5}$  \\
\rule[-2.3mm]{0mm}{5mm}
 $ B^+ \rightarrow \rho^+\rho^0$ long. & 30.1 $^{+8.3}_{-9.9}$* & 15.8 $ ^{+4.4}_{-5.2}$ \\
\rule[-2.3mm]{0mm}{5mm}
 $ B^+ \rightarrow a^+_1((\rho\pi)^+)\pi^0 $ & 35.0 $\pm$ 25.0* & 5.2 $\pm$ 3.7 \\
\rule[-2.3mm]{0mm}{5mm}
 $ B^0 \rightarrow \eta^{'}(\rho^0\gamma)\pi^0$ & 0.0 $\pm$ 1.0* & 0.0 $^{+2.0}_{-0.0}$ \\

\rule[-2.3mm]{0mm}{5mm}
 $ B^+ \rightarrow (K^{(**)}_X\pi)^+$ & 40 $\pm$ 26* & 2.3 $\pm$ 1.5 \\
\rule[-2.3mm]{0mm}{5mm}
 $ B^0 \rightarrow (K^{(**)}_X\pi)^0$ & 72 $\pm$ 54* & 3.2 $\pm$ 2.4 \\\hline
&&\\[-0.35cm]
\rule[-2.3mm]{0mm}{5mm}
 Total charmless background & - & $93.4\pm23.9$ \\ \hline
&&\\[-0.35cm]
\rule[-2.3mm]{0mm}{5mm}
 $ B^0\to X_c^0$  & - & 23.8 $\pm$ 7.1 \\
\rule[-2.3mm]{0mm}{5mm}
 $ B^+\to X_c^+$ & - & 35.2 $\pm$ 10.6  \\\hline
&&\\[-0.35cm]
\rule[-2.3mm]{0mm}{5mm}
  Total \B background &  & $152 \pm 27$ \\ \hline
\hline
\end{tabular}
\end{center}
\end{table}

%% file: MLfit.tex
\section{The Likelihood}

We use unbinned extended maximum likelihood fits to extract
the $\rhopi$ event yields, the charge asymmetries, and other
parameters used to model signal and background events. 
The fits minimize the quantity $-2\ln{\cal L}$, where ${\cal L}$ 
is the total likelihood defined over all tagging categories $\cat$ 
by
\begin{equation}
	{\cal L} = \prod_{\cat} e^{-N^{\prime}_\cat}\prod_{i=1}^{N_\cat} 
	{\cal L}_{i,\,\cat}~,
\end{equation}
and where $N^{\prime}_\cat$ is the sum of the signal and continuum 
yields (to be determined by the fit) and the fixed $B$-background 
yields, $N_\cat$ are the numbers of observed events in category 
$\cat$, and ${\cal L}_{i,\,\cat}$ is the likelihood computed for 
event~$i$. Note that no tagging information is used in the 
$B^+ \rightarrow \rho^0\pi^+$ fit. The data sample of each mode 
is assumed to consist of signal, continuum background 
and $B$-background components. The variables $\mes$, $\de$ and 
the \NN\ output discriminate signal from background. For 
$\Bz \to \rho^0\piz$, the variable~$\deltat$ is used to 
obtain additional background discrimination.

The likelihood ${\cal L}_{i,\,\cat}$ for event~$i$ is the sum of 
the probability density functions~(PDF) of all components, weighted 
by the expected yields for each component,
\beq
\label{eq:pdfsum}
	{\cal L}_{i,\,\cat} = 
		N^{\rho\pi} \epsilon_\cat {\cal P}_{i,\,\cat}^{\rho\pi}
		+ N_\cat^{q\bar{q}} {\cal P}_{i,\,\cat}^{q\bar{q}}
		 + \sum_{j=1}^{N_B}{\cal L}^{\B}_{ij,\,\cat} ~ ,
\eeq
where
\begin{itemize}

 \item 	$N^{\rho\pi}$ is the number of signal events in 
	the entire sample. For the $\rho^{+}\pi^0$ and $\rho^0\pi^{+}$ modes,
        the charge asymmetries are introduced by multiplying the signal
	yields by $\frac{1}{2}(1 - Q_B A_{CP})$, where 
	$Q_B$ is the charge of \B candidate.  

 \item 	$\epsilon_\cat$ is the fraction of signal events that are 
       	tagged in category $k$.

 \item 	$N^{q\bar{q}}_\cat$ is the expected number of continuum 
	background events that are tagged in category~$\cat$.

 \item 	The PDFs 
	${\cal P}_{\cat}^{\rho\pi}$,  
	${\cal P}_{\cat}^{q\bar q}$ and the likelihood  terms 
	${\cal L}^{\B}_{j,\cat}$ are the product of the PDFs of the 
	discriminating variables. The signal PDFs are thus given by
	${\cal P}^{(\rho\pi)^+}_\cat \equiv {\cal P}^{(\rho\pi)+}(\mes)
	\cdot {\cal P}^{(\rho\pi)+} (\de)\cdot {\cal P}^{(\rho\pi)+}_\cat (\NN)$ 
	for the charged \B decay modes, and by
	${\cal P}^{\rho^0\pi^0}_\cat \equiv {\cal P}^{\rho^0\pi^0}(\mes)
	\cdot {\cal P}^{\rho^0\pi^0} (\de)\cdot {\cal P}^{\rho^0\pi^0}_\cat (\NN)
	\cdot {\cal P}^{\rho^0\pi^0}_\cat (\deltat)$ for $\Bz \to \rho^0\piz$. 
	The PDF for continuum events is denoted 
	${\cal P}_{i,\,\cat}^{q\bar{q}}$. The likelihood term 
	${\cal L}^{\B}_{ij,\,\cat}$ corresponds to 
	the $j_{th}$ \BB\ background's contribution of the $N_B$ \B
	background categories. Correlations between the variables are
        usually neglected except that for \B backgrounds we use two-dimensional
        PDFs for $\mes$ and $\de$ to model the sizable correlations.

\end{itemize}
The signal PDFs are decomposed into two parts with distinct distributions: 
signal events that are correctly reconstructed and misreconstructed signal 
events. Moreover, for the charged \B modes we distinguish misreconstructed
signal events with right-sign $B$ charge from those with wrong-sign
$B$ charge in the likelihood. Their individual fractions are taken from 
MC simulation and given in Table~\ref{tab:effAll}. The $\mes$, $\de$, and 
\NN\  output PDFs for signal and $B$ background are taken from the simulation 
except for the means of the signal PDFs for $\mes$ and $\de$ in 
$\B^+ \rightarrow \rho^0\pip$, which are free to vary in the fit.

The parameterizations of PDFs are discussed in the following.
\begin{itemize}
\item {\boldmath$\mes$}.
	The distribution of correctly reconstructed signal is parametrized 
        using a Crystal Ball function~\cite{CB}.

	The continuum background is described by an ARGUS shape
        function~\cite{bib:argusshape} with floating shape parameter $\xi$.

\item	{\boldmath$\de$}.
	We use a Crystal Ball~\cite{CB} function (sum of two Gaussians) 
	for correctly reconstructed signal in $\rho^+\pi^0$ ($\rho^0\pi^+$
	and $\rho^0\pi^0$), and the sum of two Gaussians for misreconstructed 
	signal events. 

	Continuum background is modeled by a linear function with a 
	slope that is free to vary in the fit.

\item	{\bf \boldmath$\NN$ output}.
	PDFs for correctly reconstructed and for 
        misreconstructed signal events are taken from MC simulation
	and parameterized using empirical shape-fitting 
	techniques~\cite{keysPdfs}.
	A small discrepancy between data and MC is observed for
	the NN output distributions of control samples using 
        fully-reconstructed $B^{0} \rightarrow D^{-} \rho^{+}$ decays.
        This is accounted for in the systematic error evaluation.

        For $\bchtorchp$ and $\bchtorpch$ ($\btorp$), the PDFs describing 
        the NN output for continuum events are parametrized 
        by a third-order (fifth-order) polynomial with its parameters
   	determined in the fit.

\item 	{\bf \boldmath$\dt$}
	is used in the $B \rightarrow \rho^0\pi^0$ fit to
	improve the discrimination against continuum background.
        The distributions for correctly reconstructed signal, 
        misreconstructed signal and $B$ background events
        are treated as decays of neutral or charged \B's,
	convoluted with a $\dt$ resolution function which is 
        the sum of three Gaussians
	with parameters determined from a fit to fully 
	reconstructed \B decays~\cite{babarsin2b}. This treatment
        does not introduce a bias in the signal yield according to
        MC studies.

	The continuum $\deltat$ distribution is parameterized 
	as the sum of three Gaussian distributions with common 
	mean, two relative fractions, and three distinct widths 
	that scale the $\dt$ event-by-event error, 
	yielding six free parameters.

\end{itemize}

The shapes of the $B$ background PDFs are obtained from MC 
simulation and parameterized using empirical shape-fitting 
techniques~\cite{keysPdfs}.

We perform fits on large MC samples with the measured proportions of 
signals and continuum and $B$ backgrounds. Biases observed 
in these tests are due to imperfections in the likelihood model, {\em e.g.},
unaccounted correlations between the discriminating variables of the signal
and \B background PDFs. The observed signal yields are corrected for these 
fit biases and the full correction is assigned as a systematic uncertainty.

%% file: Results.tex
\section{Preliminary Results}
\label{sec:Results}
\begin{figure}[t]
\centerline{\epsfxsize8.0cm\epsffile{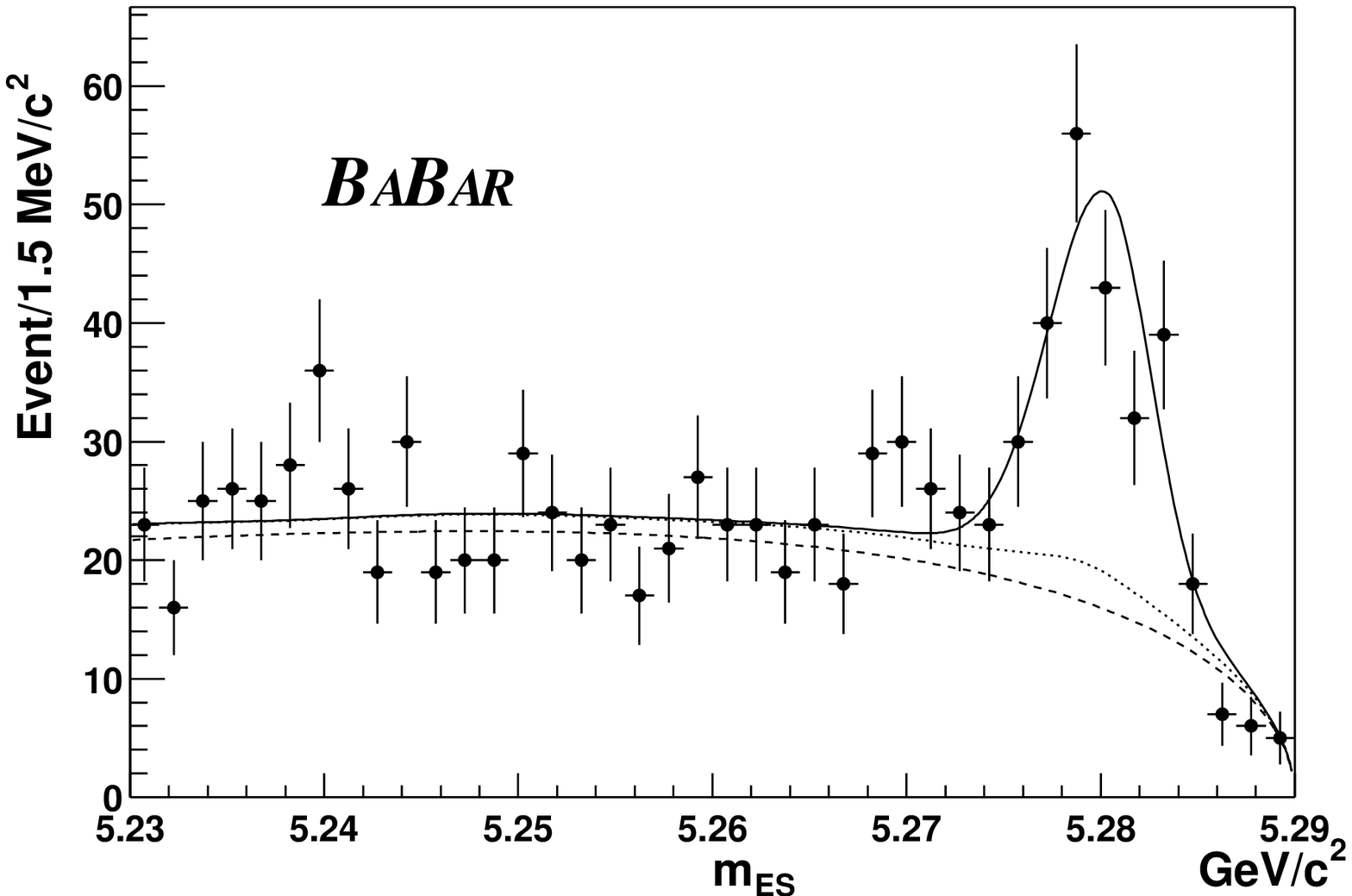}\hspace{0.3cm}
              \epsfxsize8.0cm\epsffile{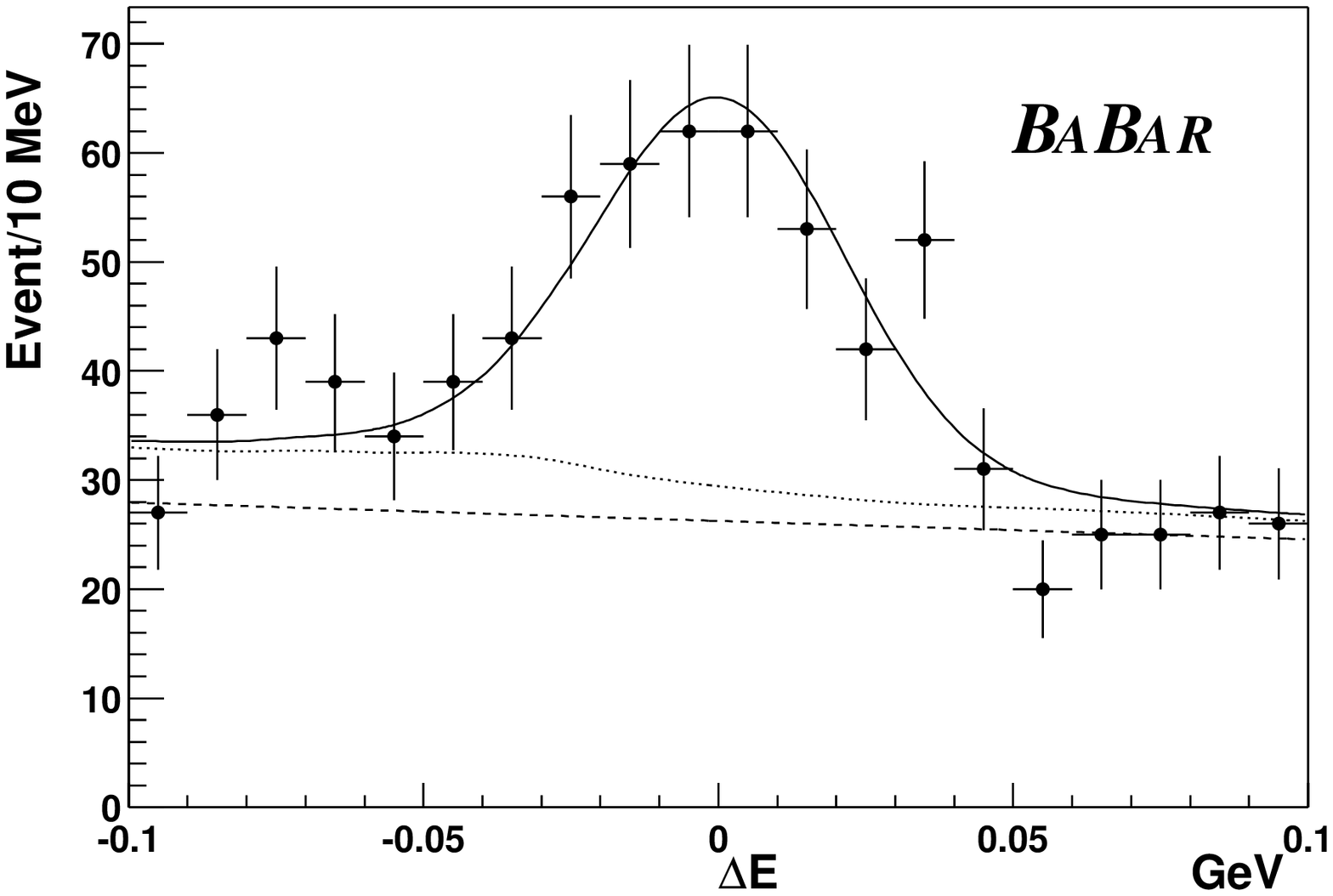}}
  \vspace{0.3cm}
\centerline{\epsfxsize8.0cm\epsffile{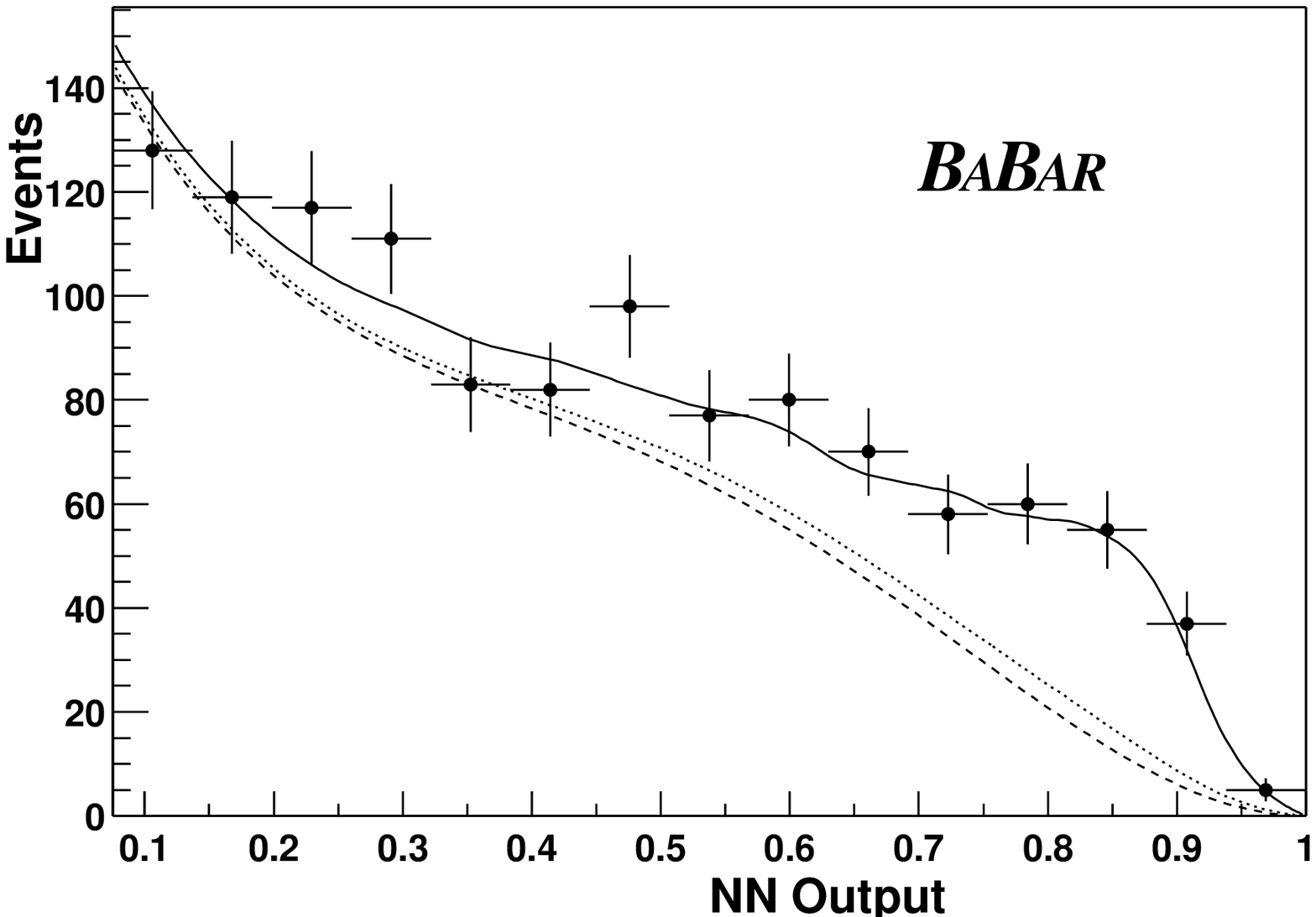}\hspace{0.3cm}
              \epsfxsize8.0cm\epsffile{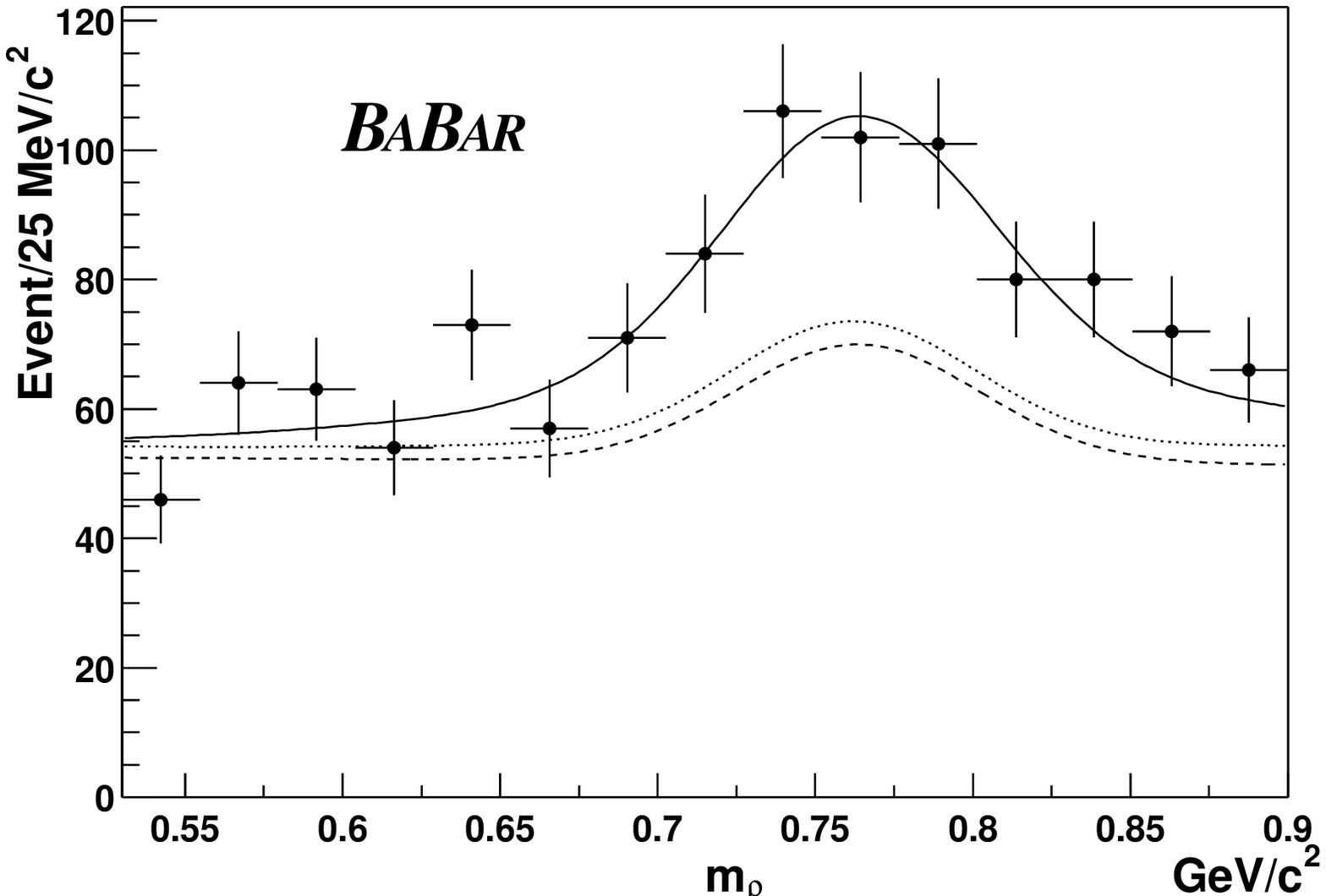}}
  \caption{\em Distributions of $\mes$ (upper left), 
	$\de$ (upper right), NN output (lower left) and the $\rho$ mass 
	(lower right) for samples enhanced in $\rho^0\pi^{+}$ signal content
	using cuts on the signal-to-continuum likelihood ratio. The 
	solid curves represent projections of the fit result. The 
	dashed curves represent the contribution from continuum events, 
	and the dotted lines indicate the combined contributions from 
	continuum events and \B backgrounds. For the $\rho$ mass distribution, 
	the fit has been repeated with $\rho$-related information 
	removed from the NN.}
\label{fig:ProjMesDE}
\end{figure}
\begin{figure}[p]
\centerline{\epsfxsize8.0cm\epsffile{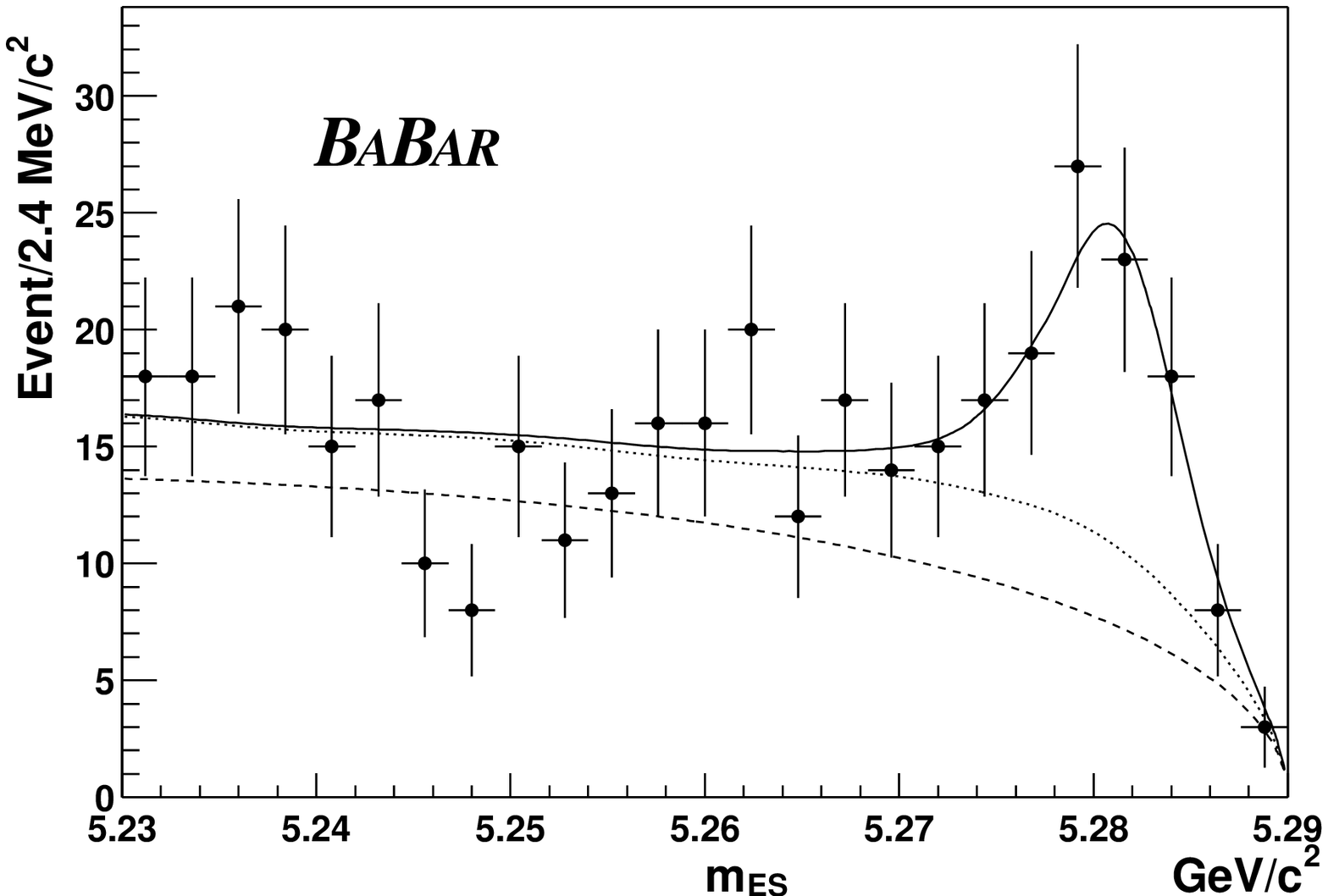}\hspace{0.3cm}
              \epsfxsize8.0cm\epsffile{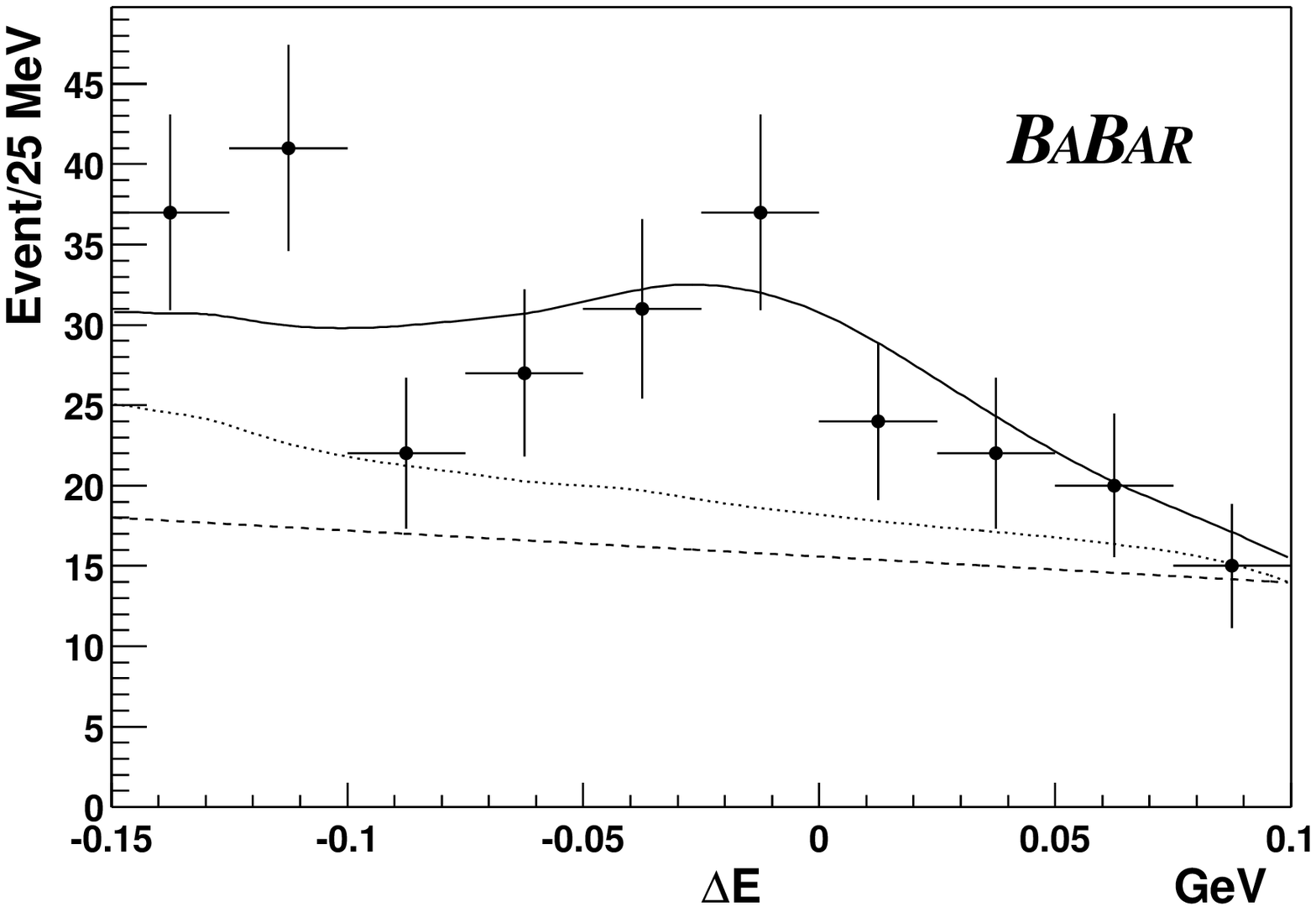}}
  \caption{\em Distributions of $\mes$ (left) and $\de$ (right)
	for samples enhanced in $\rho^+\pi^{0}$ signal content
	using cuts on the signal-to-continuum likelihood ratio. The 
	solid curves represent projections of the fit result. The 
	dashed curves represent the contribution from continuum events, 
	and the dotted lines indicate the combined contributions from 
	continuum events and \B backgrounds. }
\label{fig:ProjA}
\end{figure}

\begin{figure}[p]
\centerline{\epsfxsize8.0cm\epsffile{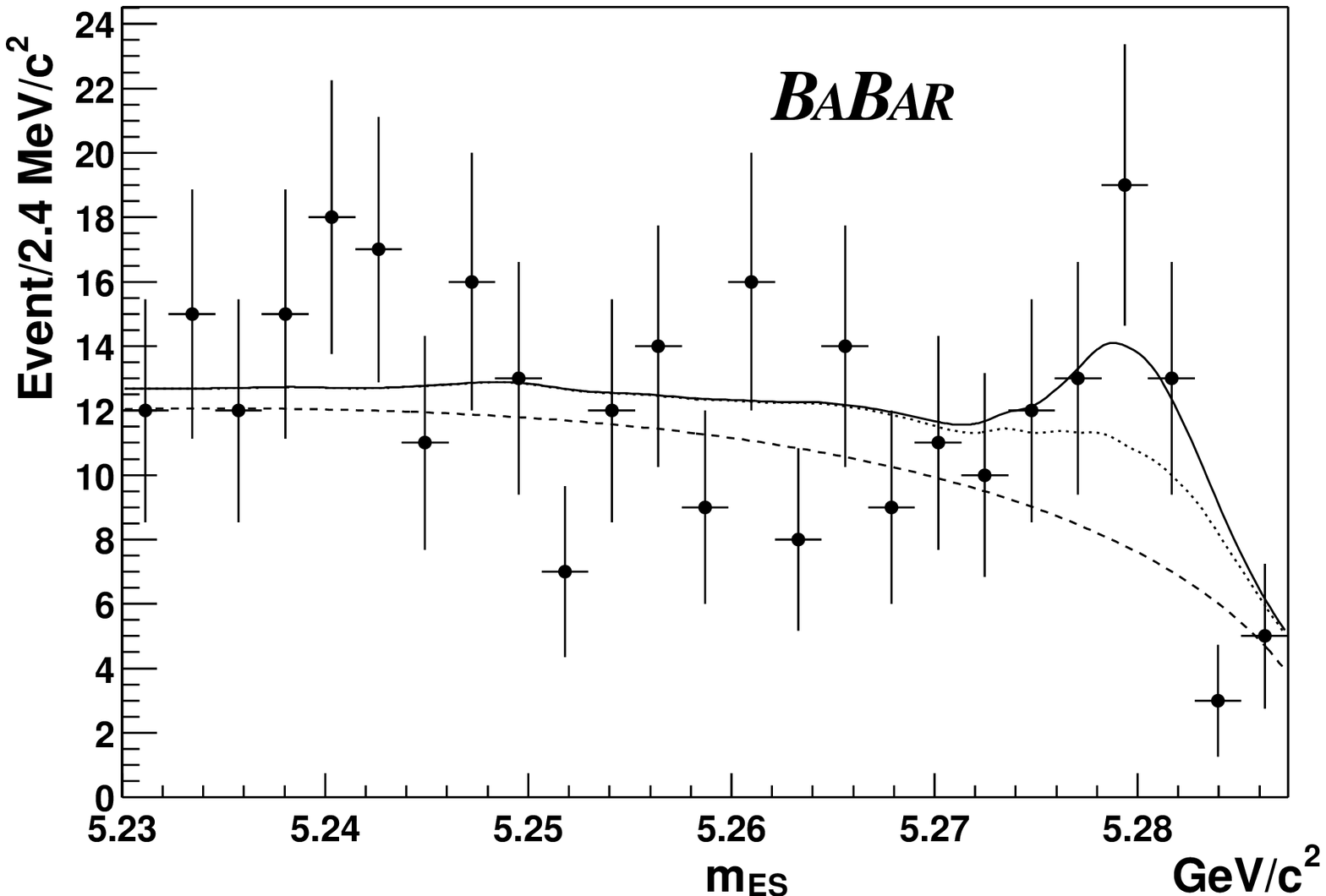}\hspace{0.3cm}
              \epsfxsize8.0cm\epsffile{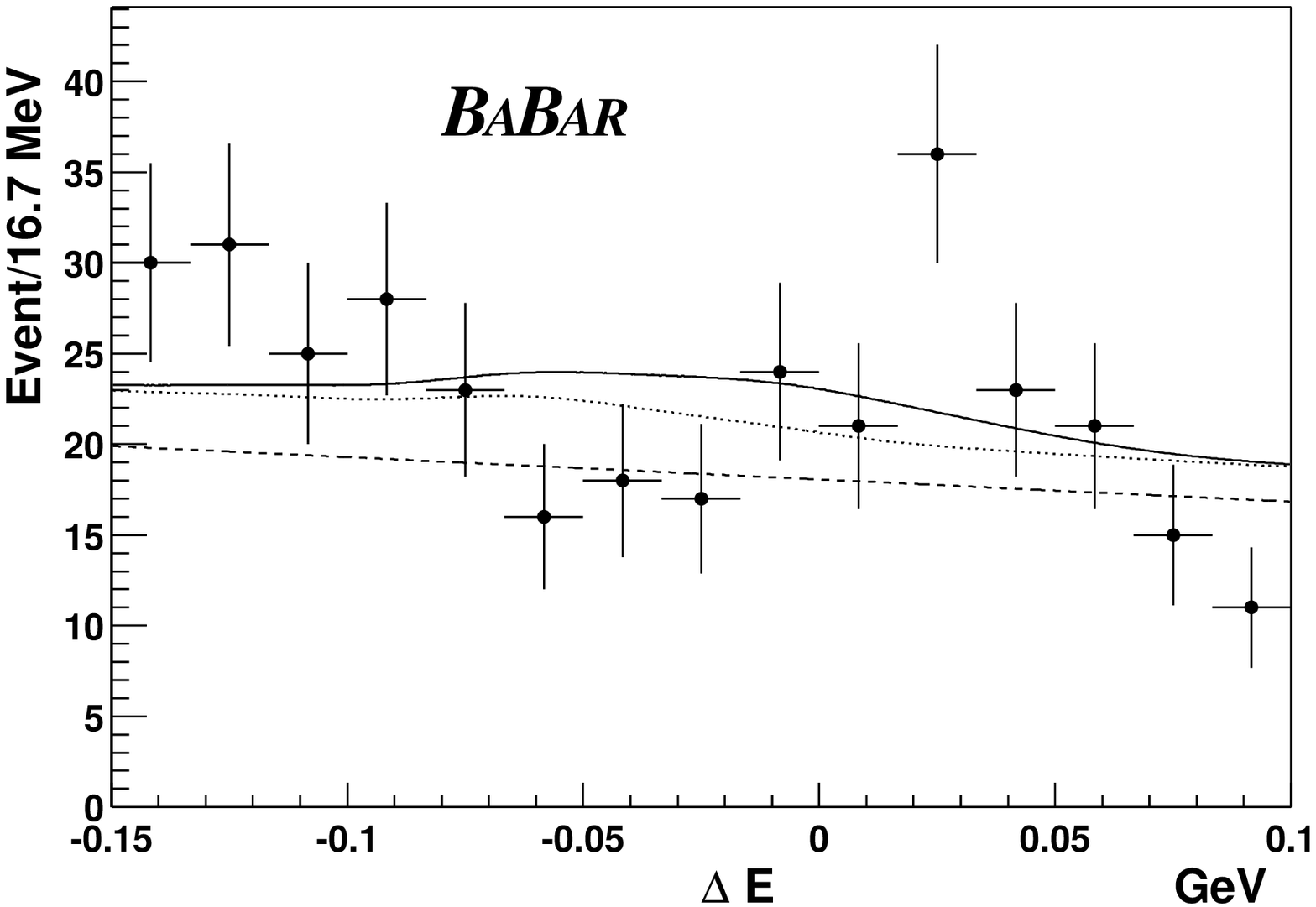}}
  \caption{\em Distributions of $\mes$ (left) and $\de$ (right)
	for samples enhanced in $\rho^0\pi^{0}$ signal content
	using cuts on the signal-to-continuum likelihood ratio. The 
	solid curves represent projections of the fit result. The 
	dashed curves represent the contribution from continuum events, 
	and the dotted lines indicate the combined contributions from 
	continuum events and \B backgrounds. }
\label{fig:ProjRho0Pi0}
\end{figure}

We obtain the event yields
$170.8 \pm 28.7{\rm (stat.)}$ for $\rho^{+}\pi^0$, 
$232.5 \pm 26.4{\rm (stat.)}$ for $\rho^0\pi^{+}$ and
$15.6 \pm 11.7{\rm (stat.)}$ for $\rho^0\pi^0$.
Assuming equal branching fractions for \FourS decays into neutral and 
charged $B$ mesons, the yields translate into the branching fractions
\beqns
{\cal B}( \B^{+}\rightarrow\rho^{+}\pi^0) 
	&=& (11.0 \pm 1.9~{\rm (stat.)}\pm 1.9~{\rm (syst.)})\times10^{-6}~, \\
{\cal B}( \B^{+}\rightarrow\rho^0\pi^{+}) 
	&=& (9.3 \pm 1.0~{\rm (stat.)} \pm 0.8~{\rm (syst.)})\times 10^{-6}~, \\
{\cal B}( \B^0\rightarrow\rho^0\pi^0) 
	&<& 2.5\times 10^{-6} \textrm{ at } 90\% \textrm{ C.L.}~,
\eeqns
where the first errors are statistical and the second systematic. 
We find the charge asymmetries:
\beqns
 	A^{\rho^{+}\pi^0}_{CP} &=& 
		0.23 \pm 0.16~{\rm (stat.)} \pm 0.06~{\rm (syst.)}, \\
	A^{\rho^{0}\pi^{+}}_{CP} &=& 
		-0.17 \pm 0.11~{\rm (stat.)} \pm0.02~{\rm (syst.)}.
\eeqns
Figure~\ref{fig:ProjMesDE} shows distributions of $\mes$, $\de$, 
the NN output and the $\rho$ mass for $\bchtorpch$, enhanced in 
signal content by cuts on the signal-to-continuum likelihood ratios 
of the other discriminating variables. The plots of $\mes$, $\de$ 
and NN correspond to the fit reported here, while the plot of 
the $\rho$ mass is obtained from a fit with $\rho$-related
information removed from the NN.

The statistical significance of the previously unobserved $\bchtorchp$ signal
amounts to $9.4\sigma$, which reduces to $6.6\sigma$ when also considering
systematic errors. Figure~\ref{fig:ProjA} shows the corresponding
signal-enhanced distributions of $\mes$ and $\de$.

For $B^0 \rightarrow \rho^0 \pi^0$, a $90\%$ confidence-level upper 
limit of 33.2 is obtained on the signal yield using a limit setting 
procedure similar to Ref.~\cite{Frequentist}. To obtain an upper limit 
for the branching ratio, the upper limit on the signal yield is shifted 
upwards by one sigma of the systematic error on the yield, and 
the efficiency and other scaling factors are shifted downwards by 
one sigma of their systematic errors. Figure~\ref{fig:ProjRho0Pi0} shows 
the corresponding signal-enhanced distributions of $\mes$ and $\de$.

All results are given in Tables~\ref{tab:yieldsum} and \ref{tab:acpsum} 
together with the systematic uncertainties discussed below.

%% file: Systematics.tex
\section{Systematic Uncertainties}
\label{sec:Systematics}

The systematic errors in the branching fractions are obtained by adding
in quadrature the systematic uncertainties in the signal yields,
the systematic uncertainties in efficiencies of tracking, particle 
identification, $\pi^0$ reconstruction and the systematic
uncertainties on the selection cuts.
The systematic errors on the $A_{CP}$ measurements
are introduced by the uncertainties in the treatment of
the \B background and by possible charge biases of
the detector.

The basis for evaluating the systematic uncertainties on the cuts that are
applied in the selection process is the differences in $\de$, $\mes$
and NN between on-resonance data and Monte Carlo simulation.
The differences between data and Monte Carlo distributions
of $\de$ and $\mes$ are extracted from various fully-reconstructed \B
control samples for the three decay modes.  The number of DIRC photons cut
for the bachelor track in $\bchtorpch$ decay mode will cause 1.0\%
uncertainty on the signal yield. The corrections
and uncertainties on the signal efficiencies are summarized in
Table~\ref{tab:yieldsum} for the three decay modes.

\begin{table}[pht]
\begin{center}
\caption[Breakdown of systematic errors for the branching ratios measurements.]
	{\em Results and breakdown of systematic errors for the branching 
	ratios measurements.}
\label{tab:yieldsum}
\vspace{0.3cm}
\begin{tabular}{lccc} \hline\hline
  &\multicolumn{3}{c}{}\\[-0.35cm]
\rule[-2.3mm]{0mm}{5mm}                      
                       & \multicolumn{3}{c}{Signal yields and efficiencies} \\ 
&&& \\[-0.35cm]
\rule[-2.3mm]{0mm}{5mm}
 & $B^+\to\rho^+\pi^0$ & $B^{+}\rightarrow \rho^0\pi^{+}$ & $B^0\to\rho^0\pi^0$ \\ \hline	
&&& \\[-0.35cm]
\rule[-2.3mm]{0mm}{5mm}
Corrected signal yield            	    & $170.8$ 		& 232.5	      & 15.6 \\
\rule[-2.3mm]{0mm}{5mm}
Corrected $\epsilon_{signal}$       & $17.5\pm0.1\%$  	&$28.3\pm0.1\%$ & $20.0\pm0.1\%$ \\ \hline
&&& \\[-0.35cm]
\rule[-2.3mm]{0mm}{5mm}
Statistical error on signal yield  & $28.7$  		& 26.4 	       & 11.7		 \\ \hline
  &\multicolumn{3}{c}{}\\[-0.35cm]
\rule[-2.3mm]{0mm}{5mm}
                       & \multicolumn{3}{c}{Yield systematics(absolute)}   \\ 
&&&\\[-0.35cm]
\rule[-2.3mm]{0mm}{5mm}
$\tau_{B}\pm 0.016 \ps$      	    & n/a		& n/a  & $0.1$   \\
\rule[-2.3mm]{0mm}{5mm}
$\Delta t$ resolution model         & n/a		& n/a  & $0.3$   \\
\rule[-2.3mm]{0mm}{5mm}
$B$ tagging                         & $3.7$  		& n/a  & $0.9$ \\
\rule[-2.3mm]{0mm}{5mm}
Fraction of misreconstructed signal  & $3.0$  		& 1.3  & $0.1$ \\
\rule[-2.3mm]{0mm}{5mm}
$\de$ PDF            	    & $8.5$ 		& 0.7  & $1.0$\\
\rule[-2.3mm]{0mm}{5mm}
$\mes$ PDF                      & $2.5$ 		& 1.6  & $1.5$ \\
\rule[-2.3mm]{0mm}{5mm}
NN PDF               	    & $3.0$ 		& 3.2  & $2.6$ \\
\rule[-2.3mm]{0mm}{5mm}
$B$ backgrounds             	    & $11.2$		& 2.3  & $^{+3.2}_{-3.9}$  \\
\rule[-2.3mm]{0mm}{5mm}
Fitting procedure         	    & $14.4$ 		& 8.2  & $6.3$ \\ \hline
&&& \\[-0.35cm]
\rule[-2.3mm]{0mm}{5mm}
Sub-total (absolute)                & $21.0$  		& 9.3  & $8.1$\\ \hline 
 & \multicolumn{3}{c}{}\\[-0.35cm]
\rule[-2.3mm]{0mm}{5mm}            
 & \multicolumn{3}{c}{Relative efficiency and scaling systematics}   \\ 
&&&\\[-0.35cm]
\rule[-2.3mm]{0mm}{5mm}
Tracking efficiency correction & $0.8\%$  			& $2.4\%$ & $1.6\%$ \\
\rule[-2.3mm]{0mm}{5mm}
PID for tracks         & $1.7 $\%			& $5.2\%$ & $4.0\%$ \\
\rule[-2.3mm]{0mm}{5mm}
Neutral correction     & $10.2\%$			& n/a 	  & $5.1\%$ \\
\rule[-2.3mm]{0mm}{5mm}
$\Delta E$ cut efficiency        & $2.6\%$ 			& $1.0\%$ & $0.1\%$ \\
\rule[-2.3mm]{0mm}{5mm}
$m_{ES}$   cut efficiency         & $0.0\%$ 			& $0.0\%$ & $0.0\%$ \\
\rule[-2.3mm]{0mm}{5mm}
$\rho$   cut efficiency         & $1.5\%$			& $0.0\%$ & $0.0\%$ \\
\rule[-2.3mm]{0mm}{5mm}
NN  cut efficiency            & $4.0\%$			& $4.0\%$ & $1.0\%$ \\
\rule[-2.3mm]{0mm}{5mm}
DIRC photons cut for bachelor $\pi^{+}$  & n/a  	& $1.0\%$ & n/a   \\
\rule[-2.3mm]{0mm}{5mm}
${\cal B}(\rho\to\pi^+\pi^-)$& $1.6\%$			& $1.6\%$ & $1.6\% $   \\
\rule[-2.3mm]{0mm}{5mm}
${\rm N}(\BB)$          & $1.1\%$ 			& $1.1\%$ & $1.1\%$ \\ \hline
&&&\\[-0.35cm]
\rule[-2.3mm]{0mm}{5mm}
Sub-total              & $11.6\%$ 			& $7.2\%$ & $7.0\%$ 	\\ \hline
&&&\\[-0.35cm]
\rule[-2.3mm]{0mm}{5mm}
Total systematic error    & $16.9\%$ 		& $8.3\%$  &  $52.5\%$\\ \hline \hline
&&&\\[-0.35cm]
\rule[-3.1mm]{0mm}{7mm}
\bf Branching ratio $[\times 10^{-6}]$   & $11.0 \pm 1.9\pm 1.9$ & $9.3\pm 1.0 \pm 0.8$ & 
$0.9 \pm 0.7 \pm 0.5$ \\ \hline \hline
\end{tabular}
\end{center}
\end{table}

\begin{table}[pht]
\begin{center}
\caption[Breakdown of systematic errors for the measurements of
   	charge asymmetries in $B^+\to\rho^+\pi^0$ and $B^+ \to\rho^0\pi^+$.]
        {\em Results and breakdown of systematic errors for the measurements of
   	charge asymmetries in $B^+\to\rho^+\pi^0$ and $B^+ \to\rho^0\pi^+$.}
\label{tab:acpsum}
\vspace{0.3cm}
\setlength{\tabcolsep}{1.5pc}
\begin{tabular}{lcc} \hline\hline
&&\\[-0.35cm]
\rule[-2.3mm]{0mm}{5mm}
& $B^+\to\rho^+\pi^0$ & $B^{+}\rightarrow \rho^0\pi^{+}$  \\ \hline	
&&\\[-0.35cm]
\rule[-2.3mm]{0mm}{5mm}
$A_{CP}$          		 & $0.23$   & $-0.17$      \\
\rule[-2.3mm]{0mm}{5mm}
Statistical error on $A_{CP}$    & $0.16$    & $0.11$     \\ \hline
 & \multicolumn{2}{c}{}\\[-0.35cm]
\rule[-2.3mm]{0mm}{5mm}
                       	 & \multicolumn{2}{c}{Fit systematics}   \\ 
&&\\[-0.35cm]
\rule[-2.3mm]{0mm}{5mm}
\B tagging              & $0.004$ & n/a\\
\rule[-2.3mm]{0mm}{5mm}
Fraction of misreconstructed signal & $0.003$ & 0.0012 \\
\rule[-2.3mm]{0mm}{5mm}
$\de$ PDF           & $0.034$ & 0.0004 \\
\rule[-2.3mm]{0mm}{5mm}
$\mes$   PDF           & $0.003$ & 0.0016 \\
\rule[-2.3mm]{0mm}{5mm}
NN PDF               & $0.00$   & 0.0028 \\
\rule[-2.3mm]{0mm}{5mm}
$B$ backgrounds             & $0.050$ & 0.022 \\
\rule[-2.3mm]{0mm}{5mm}
Detector charge asymmetry  & $0.01$  & 0.009  \\\hline
&&\\[-0.35cm]
\rule[-2.3mm]{0mm}{5mm}
Total systematic error                & $0.06$ & 0.02   \\ \hline \hline
&&\\[-0.35cm]
\rule[-3.1mm]{0mm}{7mm}
{\boldmath$A_{CP}$}       & $0.23 \pm 0.16\pm 0.06 $  & $-0.17 \pm 0.11\pm 0.02$ \\ \hline \hline
\end{tabular}
\end{center}
\end{table}

We evaluate the systematic uncertainties due to the signal $\mes$, $\de$
and NN PDFs with a large \B data control sample.
The small differences observed in the distribution shapes for Monte Carlo
events and
the distribution shapes obtained from the data control sample are used to
estimate the systematic uncertainty on the signal $\mes$ and $\de$ PDFs.
The uncertainties due to the estimated fractions of misreconstructed events
are
obtained from a control sample of fully-reconstructed $\B \rightarrow
D^-\rho^+$ decays as in Ref.~\cite{bib:350PRL}.
We perform fits on the large MC samples with the measured proportions of
$\rho^+\pi^0$, $\rho^0\pi^+$ and $\rho^0\pi^0$ signals, and continuum and \B
background. 
Fit biases observed in MC fits are added in quadrature and assigned as a 
systematic uncertainty of the fit procedure, referred to as 
``fitting procedure'' in Table~\ref{tab:yieldsum}.

The expected yields from the background modes are varied according to
the uncertainties in the measured or estimated branching fractions indicated
in Tables~\ref{tab:BbkgClass_rchp}, \ref{tab:BbkgClass_rpch} and
\ref{tab:BbkgClasse_rp}
for the $\bchtorchp$, $\bchtorpch$ and $\btorp$ decay modes, respectively.
Since \B background modes may exhibit direct \CP violation, the
corresponding parameters are varied within their physical ranges.
Contributions from non-resonant $B^+\rightarrow \pi^+\pi^0\pi^0$
for the $\rho^+\pi^0$ mode and $\B^+ \rightarrow \pi^+\pi^-\pi^+$
for the $\rho^0\pi^+$ mode are negligible according to
our dedicated studies. To check for these types of \B backgrounds,
a fit without $\rho^0$ mass and $\rho^0$ helicity information in its NN
training is performed, and the results are compatible with the fit results 
reported here. The systematic error on the $\rho^0\pi^0$ yield 
due to non-resonant $B^0\rightarrow \pi^+\pi^-\pi^0$ is considered
as part of the \B background one, based on 
Ref.~\cite{bib:neutralNR}.

For the $\bchtorpch$ and $\btorp$ decay modes,
systematic uncertainties due to possible interference between $\rho^0$ and
$f_0(980)$ or a broad scalar $\sigma(400-1200)$ are estimated to be small.
The orbital angular momentum for $\rho^0 \pi^0$ ($\rho^0\pi^+$) is one,
while for $f_0(980) \pi^0$ or $\sigma \pi^0$ ($f_0(980) \pi^+$ or
$\sigma \pi^+$) it is zero. 
Therefore the two wave functions are orthogonal. 
The interference term vanishes when integrated over the whole space.
As a cross check, MC samples with interference effects
are made from non-resonant $B^+ \rightarrow \pi^+\pi^-\pi^+$ and
$B^0 \rightarrow \pi^+\pi^-\pi^0$ Monte-Carlo using a reweighting technique.
The full selection is then applied. The relative phase is chosen
to maximize interference. Small effects are observed, as expected.

Table~\ref{tab:yieldsum}  summarizes the various sources contributing
to the systematic errors in the branching fractions.
The dominant systematic errors are due to the fit procedure 
(imperfection in likelihood model) and the uncertainties
in the \B background model.
Table~\ref{tab:acpsum}  summarizes the possible
sources contributing to the systematic errors in the charge asymmetries.

%% file: Summary.tex
\section{Summary}
\label{sec:Summary}

We have presented preliminary measurements of branching fractions and 
\CP-violating charge asymmetries in $\bchtorchp$ and $\bchtorpch$ decays,
and a search for the color-suppressed decay $\btorp$. The data sample
used in the analyses consists of $89\times10^6$ \BB\ pairs. We find a 
branching fraction for $\B^{+}\to\rho^0\pi^{+}$ that is consistent
with previous measurements~\cite{bib:belle_rhopi,bib:cleo_rhopi}. We 
observe a signal for $\B^{+}\rightarrow\rho^{+}\pi^0$ with a
significance of $6.6\sigma$, and set an upper limit for $\btorp$. We 
do not observe evidence for direct \CP violation.

%% file: pubboard/acknowledgements.tex
We are grateful for the 
extraordinary contributions of our \pep2\ colleagues in
achieving the excellent luminosity and machine conditions
that have made this work possible.
The success of this project also relies critically on the 
expertise and dedication of the computing organizations that 
support \babar.
The collaborating institutions wish to thank 
SLAC for its support and the kind hospitality extended to them. 
This work is supported by the
US Department of Energy
and National Science Foundation, the
Natural Sciences and Engineering Research Council (Canada),
Institute of High Energy Physics (China), the
Commissariat \`a l'Energie Atomique and
Institut National de Physique Nucl\'eaire et de Physique des Particules
(France), the
Bundesministerium f\"ur Bildung und Forschung and
Deutsche Forschungsgemeinschaft
(Germany), the
Istituto Nazionale di Fisica Nucleare (Italy),
the Foundation for Fundamental Research on Matter (The Netherlands),
the Research Council of Norway, the
Ministry of Science and Technology of the Russian Federation, and the
Particle Physics and Astronomy Research Council (United Kingdom). 
Individuals have received support from 
the A. P. Sloan Foundation, 
the Research Corporation,
and the Alexander von Humboldt Foundation.